\documentclass[aps,pre,twocolumn,showpacs,
amsmath,amssymb,floatfix]{revtex4}
\usepackage{graphicx}
\usepackage{dcolumn}
\usepackage{bm}
\begin{document}
\title{Coarse-graining a restricted solid-on-solid model}

\author{Achilleas Lazarides} 
\email{achilleas.lazarides@imperial.ac.uk}
\affiliation{Department of Mathematics, Imperial College 180 Queen's Gate,
  London SW7 2BZ, United Kingdom\\ and\\ Laboratorium
  voor Vaste-Stoffysica en Magnetisme, Katholieke Universiteit Leuven, B-3001
  Leuven, Belgium}
\begin{abstract}
  A procedure suggested by Vvedensky for obtaining continuum equations as the
  coarse-grained limit of discrete models is applied to the restricted
  solid-on-solid model with both adsorption and desorption.  Using an
  expansion of the master equation, discrete Langevin equations are derived;
  these agree quantitatively with direct simulation of the model.  From these,
  a continuum differential equation is derived, and the model is found to
  exhibit either Edwards-Wilkinson or Kardar-Parisi-Zhang exponents, as
  expected from symmetry arguments.  The coefficients of the resulting
  continuum equation remain well-defined in the coarse-grained limit.
\end{abstract}

\maketitle
\newcommand{\hght}{5.5cm}
\section{Introduction}
Driven, non-equilibrium interfaces have received much attention in recent
years. Various models have been used to describe such
systems.~\cite{hhz,barabasi}. These models may usually be assigned to one of a
small number of universality classes, each characterised by a set of scaling
exponents. To each universality class corresponds a continuum Langevin
equation; such an equation may therefore be identified for each lattice model
if the exponents are known, for example, from kinetic Monte Carlo (KMC)
simulations. This procedure, however, faces difficulties when crossover
effects are important~\cite{kotrlsmil}. To overcome this,
Vvedensky~\cite{vved03a} has suggested using a combination of an expansion of
the master equation~\cite{vk,fk,vzl} and the dynamic renormalisation group
(DRG)~\cite{fns,kpz} to coarse-grain the resulting
description, thereby directly obtaining a continuum Langevin equation in the
large-scale, long-time limit.  This program is a particular realization
of a program suggested by Anderson~\cite{pwa}.

In this paper we implement this procedure for a restricted solid-on-solid
(RSOS) model with both adsorption and desorption. The master equation is
expanded to obtain a set of discrete Langevin equations; these are then
numerically integrated and compared to direct KMC simulations of the model,
with which they are found to be in quantitative agreement.  In the up/down
symmetric case, an ad hoc procedure and the DRG both lead to the
Edwards-Wilkinson (EW) equation
\begin{equation}
\frac{\partial}{\partial t}\phi(x,t)=\nu_2 \frac{\partial^2}{\partial
  x^2}\phi(x,t)+\sqrt{D}\eta(x,t)
\label{eq:ew}
\end{equation}
as the macroscopic description of the model; here, $\eta$ is a zero-average
Gaussian noise field with unit variance. For the asymmetric case, DRG
arguments lead to the Kardar-Parisi-Zhang (KPZ) equation,
\begin{equation}
  \frac{\partial}{\partial t}\phi(x,t)=\nu_2 \frac{\partial^2}{\partial
    x^2}\phi(x,t)+\lambda_2\left(\frac{\partial}{\partial x}\phi(x,t)\right)^2
  +\sqrt{D}\eta(x,t)
\label{eq:kpz}
\end{equation}
as the coarse-grained description. This is consistent with symmetry arguments
and simulations~\cite{kk}. The coefficients of the coarse-grained continuum
equations remain well-defined in the macroscopic limit.

\section{\label{sec:model}The model}
\newcommand{\mh}{\ensuremath{\mathbf{h}}}
\newcommand{\mr}{\ensuremath{\mathbf{r}}}
\newcommand {\ud}{\ensuremath{\mathrm{d}}}

\begin{figure}
    \includegraphics[width=8cm]{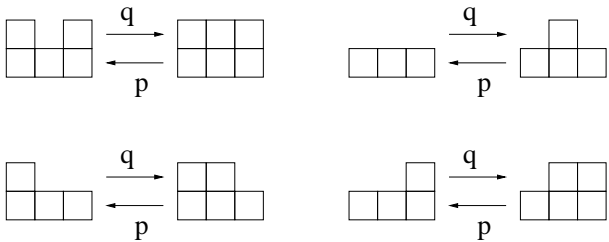}
    \caption{Allowed transitions and associated transition rates.
      \label{transitions}}
\end{figure}

The model is a simple generalisation of the RSOS model introduced by Kim and
Kosterlitz~\cite{kk}, and a special case of that introduced by Hinrichsen et
al~\cite{hlmp}. It is described (in one dimension for simplicity) by a vector
$\mh(t)$ of $L$ time-dependent, integer-valued heights $h_i(t)$,
$i=1,2,\dots,N$; time is also a discrete variable. The dynamical evolution is
given by the following rules: At each time step, an integer $1\leq i\leq L$ is
randomly chosen; with probability $p$ the height $h_i$ is increased by 1, and
with probability $q$ attempt it is decreased by 1, provided that the resulting
configuration does not violate $\vert h_i-h_{i\pm 1}\vert\leq 1$.  Finally,
periodic boundary conditions are imposed. The allowed transitions, together
with the associated probabilities (equivalently, the rates) are shown in
fig.~\ref{transitions}.
The transition rate from configuration $\mh$ to $\mh+\mr$ is given by
\begin{equation}
  \begin{split}
    W(\mh;\mr)=\sum_i\Bigg[&\Big\{q\delta(r_i
    -1)\theta(\Delta^+h_i)\theta(-\Delta^- h_i)\\
    &+p\delta(r_i
    +1)\theta(-\Delta^+h_i)\theta(\Delta^- h_i)
    \Big\}\mathcal{F}_i
    \Bigg],
    \label{transitionmatrix}
  \end{split}
\end{equation}
with $\mathcal{F}_i=\prod_{j\neq i}\delta(r_j)$, the discrete derivatives
$\Delta^\pm h_i=\pm\left(h_{i\pm1}-h_i\right)$, and
\begin{equation}
  \theta(x)=
  \begin{cases} 1, &x\geq 0\\
    0, & x<0,
  \end{cases}
  \label{thetafn}
\end{equation}
\emph{for integer (or zero) $x$}. 
The temporal evolution of the probability density $P(\mh,t)$ is given by the
master equation
\begin{equation}\begin{split}
    \frac{\partial }{\partial t}P\left(\mh,t\right)=\int \ud^L r\,
    &\Big[W\left(\mh-\mr;\mr\right)P\left(\mh-\mr,t\right)\\
    &\qquad-W\left(\mh;\mr\right)P\left(\mh,t\right)\Big].
  \label{multimeqn}
\end{split}\end{equation}

In Ref.~\cite{pk}, the special case $q=1$, $p=0$ is studied, and the KPZ
equation derived as the coarse-grained description.  However, the treatment of
Ref.~\cite{pk} leads to ill-defined coefficients precisely in the
coarse-grained limit. In particular, the authors find the coarse-grained
description to be eqn.~\eqref{eq:kpz} with $\nu_2$, $\lambda_2$ and $D$
proportional to different powers of a parameter, $a$, which controls the
degree of coarse-graining; indeed, their results are valid as $a\rightarrow
0$. Such a problem does not arise in our approach.

\section{\label{sec:variant-van-kampen}A variant of the van Kampen expansion}

An often-used method for deriving Fokker-Planck equations for stochastic
processes is the van Kampen expansion~\cite{vk,fk,vzl}. The method as
described in~\cite{vk,fk} is not directly applicable to systems such as the
RSOS model because it requires a small parameter, $1/\Omega$, in which to
expand. In effect, it \emph{assumes} the existence of a macroscopic law along
with stochastic corrections to it, the relative size of which is controlled by
the expansion parameter~\cite{vk}.  In stochastic growth models, this is not
the case; it is impossible to separate the time evolution into deterministic
and stochastic parts: the stochastic evolution is all there is. A more
extended discussion of this point may be found in
Refs.~\cite{thesis,christophalvin}, where it is shown that
eqn.~\eqref{multimeqn} may nevertheless be approximated by a Fokker-Planck
equation, which corresponds to the It\^o Langevin equation
\begin{equation}
  \frac{\ud}{\ud
    t}h_i=K^{(1)}_{i}(\mh)+\sqrt{K^{(2)}_{ij}(\mh)}\,\eta_j(t)
  \label{derSDE}
\end{equation}
where summation over repeated indices is implied and the jump moments are
defined by
\[
K^{(n)}_{i_1\dotsi_n}(\mh)=\int\ud\mr\,r_{i_1}\dots r_{i_n}W(\mh;\mr).
\]

Eqn.~(\ref{derSDE}) is essentially a set of simultaneous coupled Langevin
equations. Obtaining a continuum version is conceptually similar to the
(inverse of the) method of lines used to solve partial differential equations.
However, in the case of discrete interface models, the jump moments $K^{(n)}$
contain non-analytic step functions which necessitate a regularisation
procedure. This is done in the next section.

\section{\label{sec:discrete-langevin}Discrete Langevin equations}
To apply eqn.~\eqref{derSDE}, the first two jump moments are needed. These can
be calculated from $W$ given by eqn.~\eqref{transitionmatrix} and are
\begin{subequations}
  \begin{align}
    K^{(1)}_i(\mh)&=
    q\,\theta(\Delta^+h_i)\theta(-\Delta^- h_i)
    -p\,\theta(-\Delta^+h_i)\theta(\Delta^- h_i)
    \label{eq:k1}\\
    \intertext{and}
    K^{(2)}_{ij}(\mh)&=\left[q\,\theta(\Delta^+h_i)\theta(-\Delta^- h_i)
      +p\,\theta(-\Delta^+h_i)\theta(\Delta^- h_i)\right]\delta_{ij},
    \label{eq:k2}
  \end{align}
\end{subequations}
where $\delta_{ij}$ is a Kronecker delta.

Eqns.~\eqref{eq:k1} and~\eqref{eq:k2} are not completely specified by the
lattice transition rules of sec.~\ref{sec:model}, but must be extended to
non-integer values of the $h_i$. Furthermore, in deriving eqn.~\eqref{derSDE},
the implicit assumption that $W(\mh;\mr)$ is analytic in $\mh$ was made, which
in turn implies that the extension of $\theta(x)$ to non-integers must be
differentiable. This continuation of the $K^{(n)}$ to noninteger arguments is
known as regularisation~\cite{vzl,nagatani,vved03a,vved03b}. We choose the
representation of $\theta$ given by
\begin{equation}
  \theta(x)=\lim_{\Delta\rightarrow 0^+}\theta_{\Delta}(x),
  \label{eq:theta}
\end{equation}
where, following~\cite{nagatani,vved03a,vved03b},
\begin{equation}
  \theta_\Delta(x)=\frac{\Delta}{a} \ln\left[
    \frac{\exp\left((x+a)/\Delta\right)+1}
    {\exp\left(x/\Delta\right)+1}
  \right].
  \label{eq:thetad}
\end{equation}
The parameter $a$ is the value of $x$ below which $\theta(x)=0$ and must
therefore satisfy $0<a<1$ in order to agree with the lattice rules; it is
otherwise free at this stage. $\Delta$ is effectively a
``smoothing'' parameter, and $\theta_\Delta$ is an analytic function of $x$ at
the origin for all $\Delta>0$. More details may be found in~\cite{vved03a}.

\begin{figure}
  \begin{center}
    \includegraphics[height=\hght]{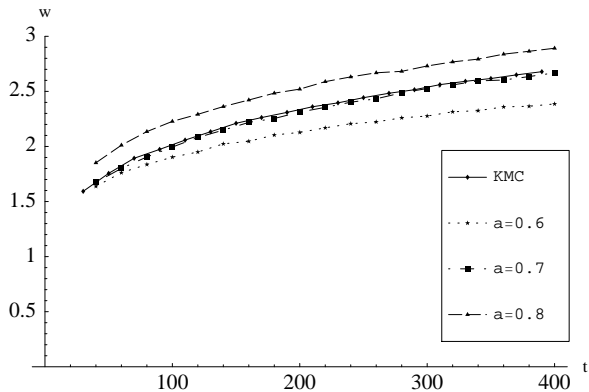}
    \caption{Variation of the evolution of the interfacial width with time for
    three values of $a$, as well as KMC simulation; $q=1$ and $p=0$, $L=100$.
    Both KMC and stochastic integration results
    averaged over 2000 runs.
      \label{fig:kmcstochcmp100}}
  \end{center}
\end{figure}

\begin{figure}
  \begin{center}
    \includegraphics[height=\hght]{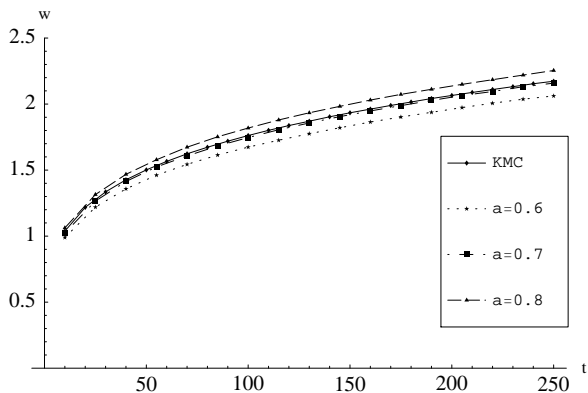}
    \caption{Early-time behaviour of the width for $L=1000$ and $q=p=0.5$ Both
      KMC and stochastic integration results averaged over 1000 runs.
      \label{fig:stochkmcew1000}}
  \end{center}
\end{figure}

In some cases, such as the Wolf-Villain model analysed
in~\cite{christophalvin}, the lattice rules can be used to infer the value of
$a$. Here, simple arguments~\cite{thesis} show that $a=0$ leads to a
discontinuous (non-analytic) dependence of $K^{(1)}(\mh)$ on $\mh$; this is
against the spirit of the model, so $a\neq 0$. Unfortunately, the argument
cannot fix the exact value; this must be determined by comparing the results
of numerical integration of eqn.~\eqref{derSDE} to simulations of the lattice
model.  Accordingly, fig.~\ref{fig:kmcstochcmp100}
shows a plot of $w$ against $t$ for $L=100$ with $q=1$ and $p=0$; results from
a KMC simulation and numerical integration of the Langevin equations for each
of $a=0.6$, $a=0.7$ and $a=0.8$ are presented. It is evident that $a\approx
0.7$ gives a result that agrees closely with the KMC simulation. To test this
agreement in the case $q=p=0.5$, we simulate a system of size $L=1000$;
fig.~\ref{fig:stochkmcew1000} shows the early-time behaviour. It again shows
that $a\approx 0.7$ gives the best agreement between KMC simulations and the
stochastic formulation.

In conclusion, the system of discrete Langevin equations with the choice
$\approx 0.7$ gives quantitatively the same results as the original
model.

\section{\label{sec:continuum-limit}The continuum limit}
The continuum (coarse-grained) limit of the symmetric case $q=p$ may be taken
using a simple ad hoc method, which is direct but physically opaque and cannot
be extended to the asymmetric case. The DRG is necessary to analyse the full
model; it is considerably more general and conceptually clear, but
correspondingly more complicated to carry out. The analysis presented in this
section confirms the results of symmetry arguments and constitutes the final
step in the identification of the SDE corresponding to the RSOS model.

\subsection{Ad hoc approach}
Using the regularisation of sec.~\ref{sec:discrete-langevin}, the step
functions may be expanded about $x=0$ as
\begin{equation}
  \label{regbn}
  \theta_{\Delta}(x)=\sum_{n=0}^\infty b_n(\Delta) x^n.
\end{equation}
Replacing the set $h_i(t)$ by a function of a continuous argument $x$,
$\phi(x,t)$ (which coincides with $h_i(t)$ for $x=i$), and expanding the
discrete derivatives transforms the discrete set of SDEs of
eqn.~\eqref{derSDE} into a (partial) stochastic differential equation for
$\phi(x,t)$.

By using power counting arguments (see~\cite{thesis} for details), we find
that the relevant terms in this equation are, at most,
\begin{widetext}
  \begin{equation}
    \label{conteqn}
    \frac{\partial\phi}{\partial t}=\nu_2 \frac{\partial^2\phi}{\partial
      x^2}+\nu_4 \frac{\partial^4\phi}{\partial
      x^4}+
    \nu_6 \frac{\partial^6\phi}{\partial
      x^6}
    +\kappa_{13}\frac{\partial\phi}{\partial
      x}\frac{\partial^3\phi}{\partial
      x^3}+\lambda_2\left(\frac{\partial\phi}{\partial
        x}\right)^2+\lambda_{13}\frac{\partial}{\partial
      x}\left(\frac{\partial\phi}{\partial
        x}\right)^3+\kappa_{22}\left(\frac{\partial^2\phi}{\partial
        x^2}\right)^2+\sqrt{D}\eta,
  \end{equation}
\end{widetext}
where $\eta$ is gaussian noise with unit variance and the coefficients may be
found explicitly as functions of the $b_n$. The coefficients $\nu_2, \nu_4,
\nu_6, \lambda_{13}$ and $D$ are proportional to $q+p$, while the rest are
proportional to $q-p$; this is consistent with arguments based on up/down
reflection symmetry. 

In the symmetric case $q=p$, the only surviving terms are given by
$\nu_2=12\nu_4=360\nu_6=b_0b_1$, $\lambda_{13}=(3 b_0b_3+b_1b_2)/3$ and $D=1$.
The coefficients $b_n$ may be found explicitly from
eqn.~\eqref{eq:thetad}. 
These satisfy the following inequality~\footnote{This may be verified by
  direct inspection of the first few coefficients; unfortunately, we have not
  found a convenient closed-form expression for $b_n$ allowing a direct
  verification. However, the conjecture is supported to at least $n=18$ by
  explicit calculation of the $b_n$~\cite{thesis}.}:
\begin{equation}
  \label{eq:conjineq}
  b_n(\Delta)\leq\frac{A_n(\Delta)}{\Delta^{n-1}},
\end{equation}
where $A_n\rightarrow\mathrm{const}$ as $\Delta\rightarrow 0^+$. 

A simple scaling approach would involve rescaling space, time and the field as
$x\rightarrow \epsilon x$, $t\rightarrow \epsilon^z t$ and
$\phi\rightarrow\epsilon^\alpha\phi$, respectively, followed by taking the
limit $\epsilon\rightarrow 0^+$.  Instead, following~\cite{vved03a,vved03b},
we will take the two limits $\epsilon\rightarrow 0$ and $\Delta\rightarrow 0$
together. Putting $\Delta=\epsilon^\delta$ with $\delta>0$ to be determined,
the limit $\epsilon\rightarrow 0^+$ is well-defined only if
$z=2$, $\alpha=1/2$ and $\delta=1/2$. The space-time scaling is diffusive
(EW), while the value of $\delta$ has no direct physical significance. At the
limit, the only surviving terms are the $\nu_2$ term and the noise; all the
others vanish. We therefore conclude that the coarse-grained limit of the
symmetric RSOS model is the EW equation~\footnote{It is important that the
  value of $\delta$ is uniquely fixed, and that the resulting coarse-grained
  equation does not depend on this value. Another crucial point is that the
  terms neglected in eqn.~\eqref{conteqn} may also be shown to vanish in the
  limit. For a detailed discussion of these points, see~\cite{thesis}.}.

Unfortunately, this direct coarse-graining procedure cannot be applied to the
asymmetric case. In the next section, both the symmetric and asymmetric
version of the model are studied using DRG arguments.

\subsection{Dynamic renormalisation group}
An equation corresponding to the limit of eqn.~(\ref{conteqn}) for $q=p$ has
previously been analysed by das Sarma and Kotlyar~\cite{dsk}; however, the DRG
flow equation for $\lambda_{13}$ is not derived in Ref.~\cite{dsk}.

For this section, eqn.~(\ref{conteqn}) will be generalised to $d$ dimensions
in the obvious way. Defining $K_d=S_d/(2\pi)^d$,
$S_d=2\pi^{d/2}/\Gamma\left(d/2\right)$ and $g=\lambda_{13}D/ \nu_2^2$, we
find, to first order in $g$ (which plays the role of an effective coupling
constant)
\begin{subequations}
  \begin{align}
    \frac{\ud\nu_2}{\ud l}&=
    \nu_2\left(z-2+\frac{1}{2}g K_d\frac{d+2}{d}\right),\\
    \frac{\ud\lambda_{13}}{\ud l}&=
    \lambda_{13}\left(2\alpha+z-4-\frac{1}{2}g
      K_d\frac{d^2+6d+20}{d(d+2)}\right),\\
    \frac{\ud D}{\ud l}&=D\left(z-d-2\alpha\right)
  \end{align}
\end{subequations}
(the non-renormalisation of $D$ is a well-known consequence of the fact that
the deterministic part of the equation is conservative). The flow equation for
$g$ is given by
\begin{equation}
  \label{sym:gflow}
  \frac{\ud g}{\ud
    l}=-g\left(d+g\frac{K_d}{2d(d+2)}\left(3d^2+14d+28\right)\right). 
\end{equation}

For any positive $g\left(l\!=\!0\right)$ (which is the case
here~\cite{thesis}), $g=0$ is an attractive fixed point~\footnote{There exists
  also a repulsive fixed point for $g<0$; however, this is irrelevant for the
  model in question because initially $g>0$. Furthermore, as pointed out to
  the author by A.~J.~Bray, had the initial $\lambda_{13}$ been negative, the
  model would not have been well defined.}. This fixed point corresponds to
$z=2$ and $\alpha=(2-d)/2$, with $\lambda_{13}=0$, that is, it is the fixed
point of the EW universality class, in agreement with the ad hoc approach of
the previous section (as well as symmetry arguments and computer simulations).

\newcommand{\mk}{\ensuremath{\mathbf{k}}}
\newcommand{\mq}{\ensuremath{\mathbf{q}}}

Application of the full machinery of the DRG to eqn.~\eqref{conteqn} is
unnecessary because definite conclusions may be reached by inspection of the
equation. In terms of the Fourier transform of $\phi(x,t)$, which we denote by
the same symbol $\phi(\mk,\omega)$, eqn.~\eqref{conteqn} (generalised to $d$
dimensions) becomes
\begin{widetext}
  \begin{equation}
    G_0(\mk,\omega)\phi(\mk)=\eta(\mk,\omega)+\int_{\mq,\Omega}
    M_2(\mq,\mk-\mq)\phi(\mq)\phi(\mk-\mq)
    +\int_{\mq,\mq'\,\Omega,\Omega'}\left[
      M_3(\mk,\mk-\mq-\mq',\mq,\mq') 
      \phi(\mq)\phi(\mq')\phi(\mk-\mq-\mq')\right].
    \label{eq:int-asym}
  \end{equation}
\end{widetext}
where the argument $\omega$ has been suppressed for $\phi$, the bare
response function is 
\[
G_0(\mk,\omega)=\nu_2 k^2-\nu_4 k^4+\nu_6 k^6-i\omega,
\]
and the two vertices are
\begin{equation}
  M_3\left(\mk_1,\mk_2,\mk_3,\mk_4\right)=\lambda_{13}
  \left[\mk_1\cdot\mk_4\right]
  \left[\mk_2\cdot\mk_3\right]
  \label{eq:3vert}
\end{equation}
and 
\begin{equation}
  \begin{split}
    M_2\left(\mk_1,\mk_2
    \right)=&-\lambda_2\left(\mk_1\cdot\mk_2\right)+\kappa_{22}
    (\mk_1)^2(\mk_2)^2 \\
    &+ \frac{1}{2}\kappa_{13} (\mk_1\cdot\mk_2)\left[(\mk_1)^2+(\mk_2)^2\right].
  \end{split}\label{eq:2vert}
\end{equation}
In the long-wavelength limit $k\rightarrow 0$ the 2-vertex of
eqn.~\eqref{eq:2vert} will be dominated by the $\lambda_2$ (KPZ) term, with
the $\kappa_{ij}$ terms playing the role of higher-order corrections.
Similarly, the 3-vertex is of higher order than the KPZ term; in addition, it
has been shown to be irrelevant previously. Therefore, the KPZ term
$\lambda_2$ determines the universality class of the model. 

Since $\lambda_2, \kappa_{ij}\propto(q-p)$, if $q=p$ then $M_2=0$ initially.
If only vertices with an odd number of legs are present in the bare
(unrenormalized) equation then vertices with an even number of vertices cannot
be produced under renormalization. Therefore, if $q=p$ the KPZ term is not
present and the coarse-grained dynamics of the system is described by the EW
equation. If $q\neq p$, the dynamics is described by the KPZ equation. This
result is consistent with symmetry arguments and simulation
results~\cite{thesis}.

\section{Summary}
We have implemented a procedure suggested by Vvedensky (and in more general
terms by Anderson~\cite{pwa}) to obtain macroscopic equations from microscopic
models. 

Discrete Langevin equations are first derived and numerically integrated; they
are found to be in quantitative agreement with KMC simulations of the
underlying model. Next, these equations are expanded leading to a continuum
Langevin equation, from which it is shown that the coarse-grained description
of the model is the KPZ equation with the coefficient of the nonlinear term
$\lambda_{2}$ vanishing in the symmetric case, so that the EW equation is
obtained. The coefficients appearing in the equation are well-defined in the
coarse-grained limit.

The advantage of this procedure over the identification of the universality
class by direct determination of the exponents from simulations is that slow
convergence to the asymptotic regime is not a problem. In addition, the DRG
approach allows, in principle, investigation of crossover effects (although
this has not been pursued here).  Application of this procedure to other
models would be a fruitful area for the future.

\begin{acknowledgments}
  The author wishes to thank A.~J.~Bray, C.~A.~Haselwandter, A.~O.~Parry and
  D.~D.~Vvedensky for useful discussions. This work was supported by a grant
  from the AG Leventis foundation.
\end{acknowledgments}


\begin{thebibliography}{199}

\bibitem{hhz} T. Halpin-Healey and Y.-C. Zhang, Phys. Rep. {\textbf 254} 215
  (1995)

\bibitem{barabasi} A.-L.~Barab\'asi and H.~E.~Stanley \emph{Fractal
    Concepts in Surface Growth} (Cambridge University Press, Cambridge, 1995)

\bibitem{kotrlsmil} M.~Kotrla and P.~Smilauer, Phys. Rev. B {\textbf 53} 13777
  (1996)

\bibitem{vved03a} D.~D.~Vvedensky, Phys. Rev. E {\textbf 67} 025102 (2003)

\bibitem{vzl} D.~D.~Vvedensky, A.~Zangwill, C.~N.~Luse and M.~R.~Wilby,
  Phys. Rev. E {\textbf 48} 852 (1993)

\bibitem{vk} N.~G.~van~Kampen, \emph{Stochastic Processes in Physics and
    Chemistry} (North Holland, Amsterdam, 1992)

\bibitem{fk} R.~F.~Fox and J.~Keizer, Phys. Rev. A {\textbf 43} 1709 (1991)

\bibitem{fns} D.~Forster, D.~R.~Nelson and M.~J.~Stephen, Phys. Rev. A {\textbf
    16} 732 (1977)

\bibitem{kpz} M.~Kardar, G.~Parisi and Y.-C.~Zhang, Phys. Rev. Lett. {\textbf
    56} 889 (1986)

\bibitem{pwa} P. W. Anderson, \emph{Basic Notions of Condensed Matter
    Physics}, Benjamin/Cummings (California, 1984), p.~212

\bibitem{kk} J.~M.~Kim and J.~M.~Kosterlitz, 
Phys. Rev. Lett {\textbf 62}, 2289 (1989)

\bibitem{hlmp} H. Hinrichsen, R. Livi, D. Mukamel and A. Politi,
  Phys. Rev. Lett. {\textbf 79} 2710 (1997)

\bibitem{pk} K.~Park and B.~Kahng, Phys. Rev. E {\textbf 51} 796 (1995)

\bibitem{thesis} A.~Lazarides, PhD Thesis, Department of Mathematics, Imperial
  College London (2005)

\bibitem{christophalvin} C. Baggio, A. L.-S. Chua, C. A. Haselwandter and
  D. D. Vvedensky, Phys. Rev. E {\textbf 72} 051103 (2005)

\bibitem{vved03b} D.~D.~Vvedensky, Phys. Rev. E {\textbf 68} 010601 (2003)

\bibitem{nagatani} T.~Nagatani, Phys. Rev. E {\textbf 58} 700 (1998)

\bibitem{dsk} S.~Das~Sarma and R.~Kotlyar, Phys. Rev. E {\textbf 50} 4275
  (1994)

\end{thebibliography}
\end{document}